\documentclass[a4paper]{article}

\usepackage{INTERSPEECH2021}
\usepackage[linesnumbered, ruled]{algorithm2e}
\usepackage{comment}
\usepackage{subfigure}
\usepackage{multirow}
\usepackage{multicol}
\usepackage{amsmath}
\usepackage{tabularx} % 导入 tabularx 包
\title{Differentiable Reward Optimization for LLM based TTS system}
\name{Changfeng Gao, Zhihao Du, Shiliang Zhang$^\dag$\thanks{$^\dag$ Work done while at Alibaba.}}
\address{Speech Team, Tongyi Lab, Alibaba Group}
\email{\{gaochangfeng.gcf, neo.dzh\}@alibaba-inc.com}
\begin{document}

\maketitle
\begin{abstract}
% This paper proposed a novel Differentiable Reward Optimization (DiffRO) to improve the performance of the neural codec language model based text-to-speech (TTS) system. Compared to other reinforcement learning from human feedback (RLHF) method for TTS, the DiffRO can directly calculate the reward from the neural codec token rather than the synthesized audio. Furthermore, we use the gumbel-softmax to make the reward function differentiable to simplify the RLHF training process. Finally, we propose a multi-task reward model to imporve the instruction following capabilities. 
% Experiments demonstrate that DiffRO can significantly improve the pronunciation accuracy of the TTS system and achieve a SOTA result on the seed-tts-eval benchmark. And with the help of the multi-task reward model, we can control the properties of emotional and quality in a zero-shot way.

This paper proposes a novel Differentiable Reward Optimization (DiffRO) method aimed at enhancing the performance of neural codec language models based text-to-speech (TTS) systems. In contrast to conventional reinforcement learning from human feedback (RLHF) approaches applied to TTS, DiffRO directly compute the rewards based on neural codec tokens, rather than relying on synthesized audio. Furthermore, we employ the Gumbel-Softmax technique to render the reward function differentiable, thereby streamlining the RLHF training process. Additionally, we introduce a multi-task reward (MTR) model which can provide feedback from different perspectives and find that it can  augment the system's capability to follow instructions effectively.
Experimental results indicate that DiffRO significantly improves the pronunciation accuracy of the TTS system, achieving state-of-the-art (SOTA) WER results on the seed-tts-eval benchmark. Moreover, with the integration of the MTR model, we demonstrate the ability to control emotional and quality attributes in a zero-shot manner.

\end{abstract}

\section{Introduction}

Inspired by the success of text-based large language models (LLMs), neural codec token language modeling (LM) has emerged as a leading approach for text-to-speech (TTS) generation. This methodology employs a pre-trained speech tokenizer to encode speech into a discrete sequence of tokens, which are subsequently modeled using a decoder-only LM that predicts these tokens based on textual input. Ultimately, a flow matching (FM)\cite{NEURIPS2020_4c5bcfec} model and a vocoder are applied to transform the tokens into audible speech. Leveraging the capabilities of zero-shot and in-context learning inherent in LLMs, neural codec-based TTS systems \cite{lmtts1, lmtts2} have demonstrated enhanced quality and naturalness, enabling applications such as zero-shot TTS and instruction-based TTS.

In the training of LLMs, Reinforcement Learning from Human Feedback (RLHF) \cite{ziegler2020finetuninglanguagemodelshuman} represents a critical step in aligning LLMs with human preferences, contributing to the success of numerous state-of-the-art models. Recent TTS systems have also endeavored to incorporate RLHF to enhance the quality of generated speech; however, a widely applicable methodology remains elusive. The challenges can be summarized as follows:

Firstly, unlike traditional natural language processing (NLP) tasks, TTS systems require additional backend  FM  and vocoder models to convert discrete neural codec tokens into waveform audio. The computational demands posed by these backend models are substantial, which hinders the large-scale production of RLHF data.

Secondly, the diversity of generated TTS speech is often insufficient. Although the neural codec LLM can generate different token sequences from the same text and speech prompt by employing various sampling strategies and random seeds, the resultant voices consistently exhibit high similarity. Consequently, it becomes challenging to differentiate between positive and negative feedback for training the reward model.

Finally, the evaluation methods for TTS systems are inherently complex. Typically, the mean opinion score (MOS) serves as the gold standard for evaluating TTS quality. However, more critical aspects such as accuracy, naturalness, and speaker similarity are ultimately determined by listener perception. This issue is exacerbated in instruction-based TTS scenarios, where certain instructions may conflict with the need for precise pronunciation accuracy.

Several existing studies \cite{tts_rlhf, chen_enhancing_2024, hu2024robustzeroshottexttospeechsynthesis, anastassiou_seed-tts_2024, gao_emo-dpo_2024} have attempted to address one or more of the aforementioned challenges through various approaches. For instance, Chen et al. \cite{chen_enhancing_2024} introduced a uncertainty-aware optimization (UNO) with  unpaired negative and positive samples which can mitigate the impact of insufficient diversity within TTS systems. They further proposed a Reverse Input Output (RIO) strategy to generate additional positive samples by reversing the prompts and their corresponding responses \cite{hu2024robustzeroshottexttospeechsynthesis}. Anastassiou et al. \cite{anastassiou_seed-tts_2024} utilized automatic speech recognition (ASR) and speaker recognition systems to measure word error rate (WER) and speaker similarity as reinforcement learning (RL) rewards. Moreover, Gao et al. \cite{gao_emo-dpo_2024} demonstrated that employing the accuracy of the speech emotion recognition (SER) model as a reward can enhance emotional controllability. Collectively, these studies illustrate that leveraging objective indicators can yield significant improvements in TTS systems across various dimensions.

In this work, we propose a novel Differentiable Reward Optimization (DiffRO) method specifically designed for neural codec-based TTS systems. Unlike existing approaches, DiffRO predicts rewards directly from the neural codec tokens rather than from the synthesized speech, thereby reducing the computational burden associated with backend frame-matching and vocoder models. Furthermore, we utilize the Gumbel-Softmax technique to ensure that the loss function of the reward model is differentiable with respect to the input text of the TTS system. Finally, we construct a multiple downstream tasks reward (MTR) model, including ASR, speech emotion recognition (SER), speech quality assessment (SQA), and age and gender prediction. Our findings indicate that the MTR model can effectively regulate speech attributes in accordance with preset preferences or input instructions.

\section{Related work}
\label{S:2}
\subsection{Neural Codec LLM based TTS systems}
% Most neural codec based LM TTS systems consists of four main building blocks: a pre-trained speech tokenizer, a neural codec token language model, a flow matching model, and an acoustic vocoder. Given a pairwise speech $X_{1:T}$ and text $Y_{1:N}$. The speech tokenizer will encode the speech into token sequence $U_{1:T}$ and then train the language model with a next token prediction tasks:

\begin{figure*}
    \centering
    \includegraphics[width=0.7\textwidth]{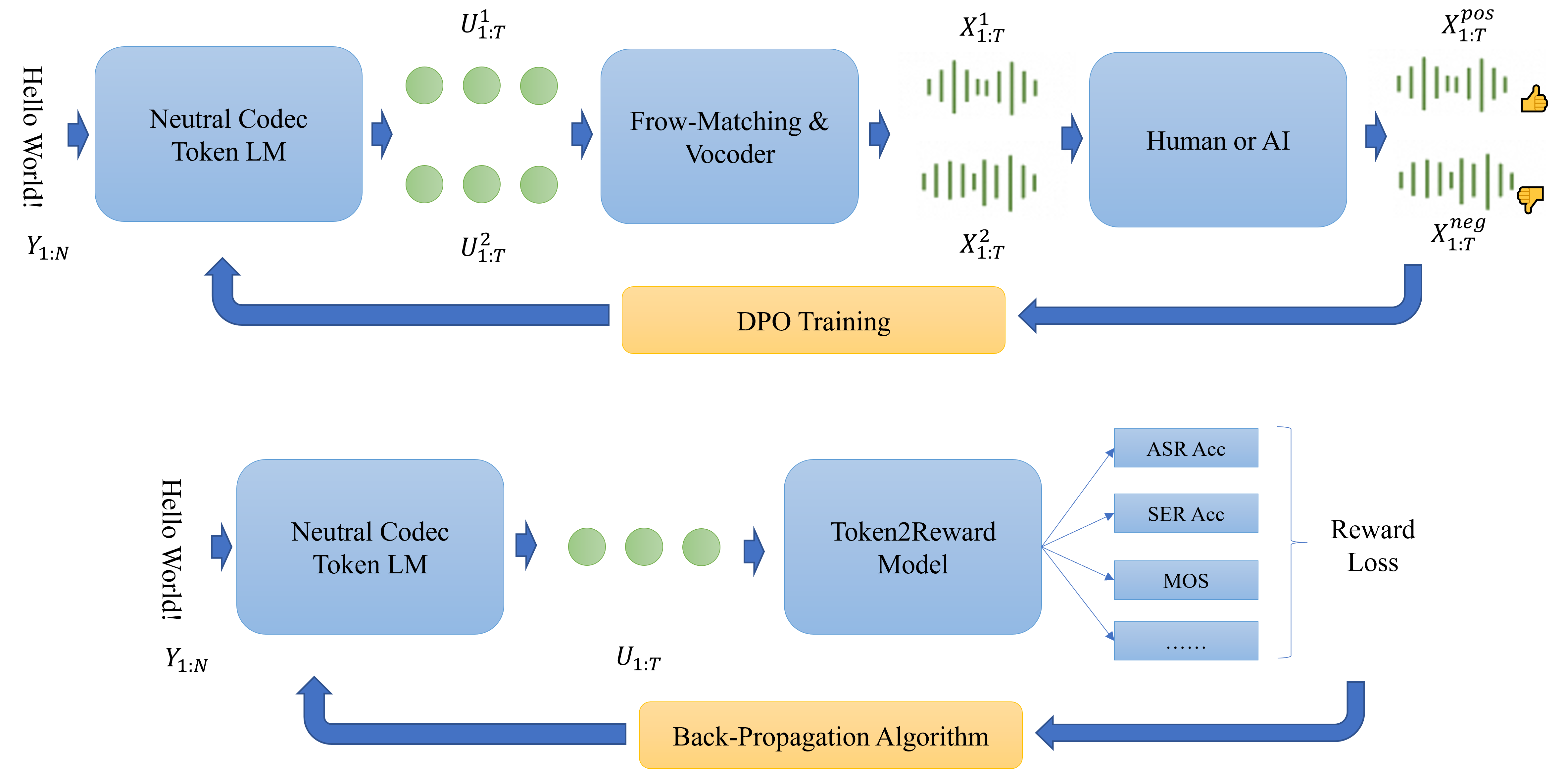}
    \caption{Comparison between the DPO an the proposed DiffRO. DiffRO can directly predict the reward score from the codec token sequence rather than the audio, and it can directly optimize the LM with back-propagation rather than the RL training loop like DPO.}
    \label{fig:DiffRO_vs_dpo}
\end{figure*}

Most neural codec LLM based TTS systems comprise four primary building blocks: a pre-trained speech tokenizer, a neural codec token LM, a FM model, and an acoustic vocoder. Given a paired input of speech  $X_{1:T}$ and text $Y_{1:N}$,  the speech tokenizer encodes the speech into a token sequence $U_{1:T}$. Subsequently, the LM is trained using a next-token prediction task:
\begin{equation}
  U_t = \text{argmax}_{1:Q} P_{\pi}(U_t = q | U_{1:t-1}, Y_{1:N})  
\end{equation}\label{eq:lm}
where $Q$ is the code-book size of the tokenizer, $\pi$ stand for the LM model.

In general, the token sequence $U_{1:T}$ carries most information of the text and plays a crucial role in the TTS system. The TTS system will recover the audios $X_{1:T}$ from $U_{1:T}$ with the FM and the vocoder.

\begin{equation}
X_{1:T} = \text{Vocoder}(\text{FM}(U_{1:T}; \theta_{f}); \theta_{v})
\end{equation}
where $\theta_{f}$ and $\theta_{v}$ are the parameters of the FM model and the vocoder.

\subsection{RLHF for TTS}
To obtain human or AI feedback, one input $Y$ should be inputted into the LM more than once, then different sample strategy and random seed are used to obtain different audios ${X^1,  X^2,...,X^K}$. The positive and negative $X^{pos}$ and $X^{neg}$ are chosen from the $\{X^i\}_{i=1}^K$ and their neural codec tokens $U^{pos}$ and $U^{neg}$ 
are used to train a reward model $R(X, Y)$, which can provide feedback to the LM during the RL phase.
% and re-encoded into the token sequence $U^{pos}$ and $U^{neg}$ for reinforcement leaning.
\begin{equation} \label{reward}
% \begin{align}
    \mathcal{L}_R= -\mathbb{E} \left[ \log \sigma(R(U^{pos}, Y) - R(U^{neg}, Y)) \right]
    % a
\end{equation}
\begin{equation} \label{rl}
    \pi_\theta^* = \max_{\pi_\theta} \mathbb{E} \left[ R(U, Y) \right] - \beta D_{\text{KL}} \left[ \pi_\theta(U|Y) \| \pi_{\text{ref}}(U|Y) \right]
% \end{align}
\end{equation}
Proximal policy optimization (PPO) \cite{schulman_proximal_2017} are used to optimize the eq \ref{rl}. Direct Preference Optimization (DPO) \cite{rafailov_direct_2023} can merge the reward modeling and RL into one phase and optimize the eq \ref{reward} and \ref{rl} with $U^{pos}$ and $U^{neg}$ directly.
\begin{equation}
\begin{split}
    \mathcal{L}_{DPO}(\pi_\theta, \pi_{ref}) =  -\mathbb{E} \left[\log \sigma \left( \beta \log \frac{\pi_\theta(U^{pos}|Y)}{\pi_{ref}(U^{pos}|Y)} \right. \right. \\ \quad \left. \left. - \beta \log \frac{\pi_\theta(U^{neg}|Y)}{\pi_{ref}(U^{neg}|Y)} \right)\right]
\end{split}
\end{equation}
where $\pi_{\theta}$ is the optimized LM model and $\pi_{ref}$ denotes a frozen reference model.

% Although the DPO can simplify the RL into one phase, however, constructing a preferred and dispreferred training set is still quite complex task for the TTS system. 
% On the one hand, to obtain the final feedback, the predicted token sequence should be decoded into the waveform by the flow-matching model and the vocoder. On the other hand, the synthesized audios could be similar with each other, it is difficult to distinguish the positive and the negative from only two samples.
% Besides the complexity, another weakness of the DPO is that simply dividing the audios into good and bad may not be suitable for the TTS. The quality of the audio should be evaluated from more than one aspect like pronunciation, rhythm, speaker similarity, emotional expression and etc. 

Although DPO can simplify RLHF into a single phase, constructing a preferred and dispreferred training set remains a complex task for TTS systems.
On one hand, to derive meaningful feedback, the predicted token sequence must be decoded into waveform audio via the FM model and the vocoder. On the other hand, synthesized audio samples often exhibit high similarity, making it challenging to accurately distinguish between positive and negative examples based solely on two samples.
In addition to this complexity, another limitation of DPO is that categorizing audio into good and bad may not be adequate for TTS. The quality of audio output should be assessed from multiple perspectives, including pronunciation, rhythm, speaker similarity, emotional expression, and other relevant attributes.

\section{Differentiable Reward Optimization}
\label{S:3}

This section will introduce the DiffRO, which can further simplify the RL training process and can provide feedback on different aspects. Figure \ref{fig:DiffRO_vs_dpo} shows the difference between the DiffRO and the existing RL method like DPO.

\subsection{Token2Reward Prediction} 

Unlike previous work, DiffRO directly predicts the reward from the speech tokens rather than the raw audio. As a TTS system should read the text correctly, the predicted codec token $\tilde{U}$ should catch all information from the text. So we can predict the input text $Y$ from the code $U$ with a neural network in ASR way \cite{10374223}:
\begin{equation}
\tilde{U}_t = \text{argmax} P_{\pi_{\theta}}(U_t | U_{1:n-1}; Y)   \label{eq:gumbel}
\end{equation} 
\begin{equation}
Y_n^* = \text{argmax} P_{ASR}(\tilde{Y}_n | Y_{1:n-1}; \tilde{U}_{1:T})   
\end{equation}

The post-probability can be regarded as the reward model, as it can encourage the $\tilde{U}$ catch more information from the text. 
We further use the Gumbel-Softmax operation to replace the argmax operation in eq \ref{eq:gumbel} to sample the predict token $\tilde{U}$. Then the reward function is differentiable and the LM model can be directly optimized to maximize the reward score without PPO or DPO strategy. 
\begin{equation}
    R_{ASR}(Y) = \log P_{ASR}(\tilde{Y}_n = Y_n | Y_{1:n-1}; \tilde{U}_{1:T})   
\end{equation}

\subsection{Multi-Task Reward Model}

% Besides the ASR reward, we can also use more downstream tasks to guarantee the predicted token sequence contains sufficient information that we need. 
% We first train a codec-based speech understanding model with multi-task training which can do SER, SQA, AED, MOS prediction, and other audio understanding tasks. Then we use the this model as a multi-task reward (MTR) model to guide the TTS system generate audios follow special instruction or have specific characteristics. The structure of the MTR model is shown in Figure \ref{fig:mtr}.

\begin{figure}
    \centering
    \includegraphics[width=0.7\linewidth]{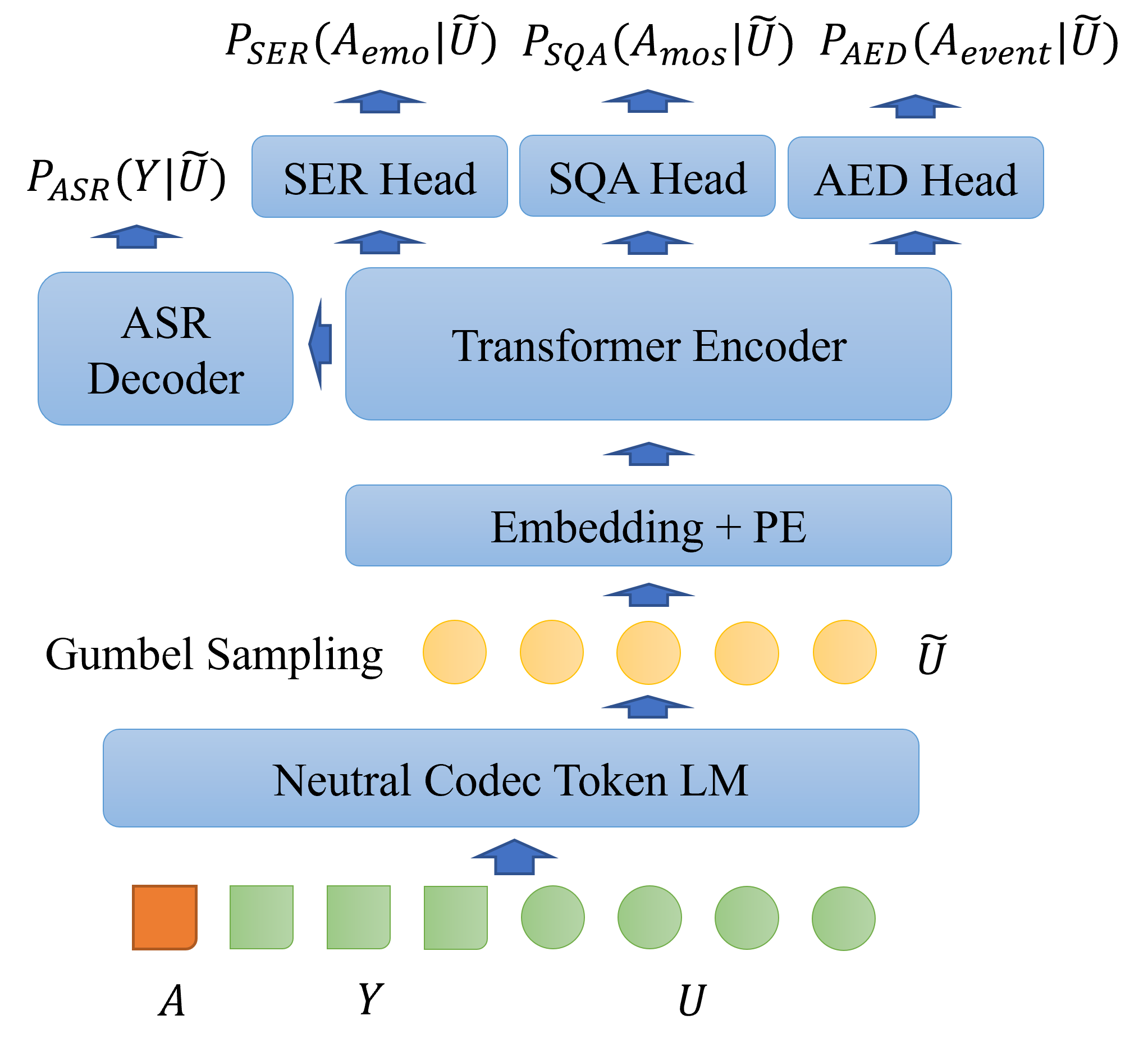}
    \caption{The Structure of the proposed MTR model. ASR, SER, SQA and AED are used for the downstream task in this figure and can be adjusted according to requirements.}
    \label{fig:mtr}
\end{figure}

In addition to utilizing ASR rewards, we can incorporate additional downstream tasks to ensure that the predicted token sequence encompasses all the requisite information.
Initially, we train a codec-based speech understanding model using a multi-task training approach, enabling it to perform tasks such as SER, SQA, AED and other audio understanding functionalities. Subsequently, we employ this model as a MTR model to guide the TTS system in generating audio that adheres to specific instructions or exhibits particular characteristics. The structure of the MTR model is illustrated in Figure \ref{fig:mtr}.

If the MTR model is accessible, we can control the synthesized audio's attribute $\{A_i\}_{i=1}^K$ or make the TTS system follow some instruction by maxmize the post-probably predicted by the MTR model.
\begin{equation}
R_{MTR}(Y, \{A_i\}_{i=1}^K) = \sum_i \log P_{\text{task}_i}(\tilde{A_i} = A_i | \tilde{U})
\end{equation}

% \begin{equation}
% \tilde{U_t} = \text{GumbelMax}_{1:Q} P_{\pi}(U_t = q | U_{1:t-1}, Y_{1:N}, A)
% \end{equation}
Where $Y$ is the text, $\tilde{U}$ is the predict token and $A$ is the target attribute and $i$ is the task id. 
% It should be noticed that if the $P_{MTR}$ is available, no audio are needed during the training of the RL. 

\section{Experiments}
\label{S:4}
\subsection{Experimental Setup.}
\subsubsection{DataSet and Baseline TTS system} 
We use the Cosyvoice2.0-0.5B \cite{du2024cosyvoice2scalablestreaming} model as the baseline system and fine-tune it with our in-home target speaker data as the SFT baseline. The SFT data contains 4000 audio samples from 5 speakers (4 female and 1 male) and all of them speak in Mandarin.
For the RL data, we collect ten thousand texts from the Internet, 90\% of which are Chinese and English, and others are Japanese and Korean. Then we use the Cosyvoice2.0 to synthesize corresponding audios.

\subsubsection{Training of the Reward Model}

For the reward model, we replace the front-end CNN module of the SenseVoice with an embedding layer to accept the speech token input and add an attention pooling layer for each downstream task. 
And for the multi-task training data, we generate pseudo labels by the existing models for the inhome industry ASR training corpus. The final data obtained contains more than 13000 hours audios, and each audio has their transcription, emotion\footnote{https://modelscope.cn/models/iic/emotion2vec\_plus\_large/}, 
MOS\cite{Kumar2023TorchaudioSquimRS}, 
speaker age\&gender\footnote{https://huggingface.co/audeering/wav2vec2-large-robust-24-ft-age-gender}, and event labels\cite{Kong2019PANNsLP}. 
The emotion and gender are predicted by cross entropy (CE) loss, the age are predicted by mean squared error (MSE) loss and the events are predicted by binary CE loss. While for MOS prediction, we divide the audio quality into 5 levels and also predict it by CE loss.

After training, we evaluate this token2reward model on the downstream tasks \cite{commonvoice:2020, huang22f_interspeech, busso2008iemocap, chen2024airbench} and show the result in Table \ref{tab:mtr_reward}. We can find that for the ASR task, the WER is a little worse than the SenseVoice\cite{an2024funaudiollmvoiceunderstandinggeneration}; this is reasonable because compared with the raw audio, the information in the codec token is incomplete. However, when it comes to the MTR, our reward model has better SQA, Age\&Gender prediction capability than the reference\cite{reddy2022dnsmos, ma2024emoboxmultilingualmulticorpusspeech, Qwen-Audio}. 
This proves that the codec tokens contain sufficient emotion, speaker and noise information, which means that the LM based TTS system has the potential to synthesize rich expression audio. 
This also proves that we can also train a multi-task speech understand model under the traditional classification framework rather than the LLM based generative framework \cite{Qwen-Audio, chen2025minmomultimodallargelanguage}.

% \begin{table}[!t]
%     \centering
%     \begin{tabular}{l ll cc}
%     \hline
%        \multirow{2}{*}{Reward} & \multirow{2}{*}{Task\&Data} &\multirow{2}{*}{Ref-Source}  & \multicolumn{2}{c}{Value}  \\
%        &  & & ref. & res.  \\
%     \hline
%         \multirow{2}{*}{ASR} &  ASR-CMV(en) &  SenseVoice  &  \textbf{8.67} &  11.3  \\
%          &  ASR-CMV(zh) &  SenseVoice  &  \textbf{7.59} &  7.70  \\
%          \hline
%          \multirow{4}{*}{MTR} &  SQA-BVCC &  DNSMOS &  0.61 &  \textbf{0.65}  \\
%          &  SER-IEMOCAP &  EmoBox       &  \textbf{72.8} &  66.0  \\
%          &  Age-AirBench &  Qwen-Audio &  58.8 &  \textbf{65.8}  \\
%          &  Gender-AirBench &  Qwen-Audio &  82.5 &  \textbf{92.5}  \\

%     \hline
%     \end{tabular}
%     \caption{Performace of the token2reward model on different downstream task . ASR, SER, SQA, Age\&Gender tasks are evalueted with WER(\%), Weighted Accuracy(\%), linear correlation coefficient and Accuracy(\%).}
%     \label{tab:mtr_reward}
% \end{table}

\begin{table}[!t]
    \centering
    \begin{tabular}{ll cc}
    \hline
       \multirow{2}{*}{Task\&Data} &\multirow{2}{*}{Ref-Source}  & \multicolumn{2}{c}{Value}  \\
        & & refer & result  \\
    \hline
         ASR-CMV(en)-WER(\%) &  SenseVoice  &  \textbf{8.67} &  11.3  \\
           ASR-CMV(zh)-WER(\%) &  SenseVoice  &  \textbf{7.59} &  7.70  \\
         \hline
           SQA-BVCC-LCC &  DNSMOS &  0.61 &  \textbf{0.65}  \\
          SER-IEMOCAP-WA(\%) &  EmoBox       &  \textbf{72.8} &  66.0  \\
          Age-AirBench-Acc(\%) &  Qwen-Audio &  58.8 &  \textbf{65.8}  \\
          Gender-AirBench-Acc(\%) &  Qwen-Audio &  82.5 &  \textbf{92.5}  \\

    \hline
    \end{tabular}
    \caption{Performace of the token2reward model on different downstream task . ASR, SER, SQA, Age\&Gender tasks are evalueted with WER(\%), Weighted Accuracy(\%), linear correlation coefficient and Accuracy(\%).}
    \label{tab:mtr_reward}
\end{table}

\subsubsection{Reinforcement Learning Setup}
For the RL, we compare the DPO and the proposed DiffRO. For the DPO, we synthesize the text five times and select the $X^{pos}$ and $X^{neg}$ according to the WER and speaker similarity. 
For DiffRO, we utilize the ASR reward to enhance pronunciation, while employing the MTR to control the audio properties. $\beta$ is set to 0.1 for all experiments and the learning rate is fixed to $0.00001$. 4 A800 GPU are used for the training. 

% \subsection{Experiments Results}

\subsection{Results for ASR-based DiffRO}

We evaluate the TTS system on the seed-tts-eval benchmark by the WER of the ASR models. For the Chinese \textit{zh} and \textit{hard} subset, we use the Paramformer-zh \cite{gao22b_interspeech} and for the \textit{en}, we use the whisper-large-v3\cite{radford2022robustspeechrecognitionlargescale}. We further also evaluate the Japanese and Korean in CV3-Eval \footnote{https://github.com/FunAudioLLM/CV3-Eval} by whisper. Results are shown in Table \ref{tab:asr_result}.

\begin{table}[!t]
    \centering
    \begin{tabular}{l ccc cc}
    \hline
        \multirow{2}{*}{Model} & \multicolumn{3}{c}{Seed-TTS}& \multicolumn{2}{c}{CMV} \\
        & zh & en & hard & ja & ko \\
    \hline
        CosyVoice2.0 &  1.56 &  2.75 &  6.91 & 9.13 & 7.43 \\
         % + DiffRO-ASR &  \textbf{0.81} & \textbf{2.50}  & \textbf{5.51}  & \textbf{8.84} & \textbf{5.69} \\
         + DiffRO-ASR &  \textbf{0.78} & \textbf{1.89}  & \textbf{5.58}  & \textbf{6.36} & \textbf{5.41} \\
    \hline
        CosyVoice2.0-SFT &  1.50 &  4.26 &  7.90 & 20.2 & 11.1 \\
         + DPO &  1.27 & 3.28 & 6.74 & 10.4 & 9.12 \\
         + DiffRO-ASR & 1.09 & \textbf{2.57} & \textbf{5.83} & \textbf{8.38} & \textbf{6.35} \\
         + DiffRO-MTR & \textbf{1.05} & 3.43 & 6.41 & 14.8 & 10.7 \\
    \hline
    \end{tabular}
    \caption{WER(\%) for audios synthesized from Different TTS system.}
    \label{tab:asr_result}
\end{table}

According to the table, we can find that after speaker fine-tuning (SFT), the TTS system makes some improvement on the \textit{zh} subset but becomes a little worse on the \textit{en} and \textit{hard}. This is reasonable as the SFT data are mostly composed of Mandarin and in lack of hard case. And the decrease in \textit{ja} and \textit{ko} could be more significant due to catastrophic forgetting.
While RL can improve the TTS performance on all sets, and the DiffRO-ASR is the best one. It can even beat the base model on \textit{ja} and \textit{ko}, although only a small amount of Japanese and Korean text can be utilized during RL. 
Because the TTS can learn to generate multilingual codec-tokens by the ASR model.
The DiffRO can also be applied to the base model and achieves state-of-the-art result for \textit{zh} and \textit{hard}. 
However, in the ASR task, the MTR-based DiffRO demonstrates inferior performance compared to the ASR-only DiffRO model. This discrepancy arises because the MTR approach encourages the predicted tokens to carry additional information that is not utilized by the ASR system. Nevertheless, this extra information can facilitate audio style control, a topic we will discuss further in later subsections.

\subsection{Results for MTR-based DiffRO}

\subsubsection{Emotion Control}

We can utilize the SER reward to the emotion of the output audio, and the data are structured according to the following instruction template:

    $\text{Your emotion is } E \text{ $\langle$endofprompt$\rangle$ } Y_{1:N} \text{ $\langle$s$\rangle$ }U_{1:T} \text{ $\langle$/s$\rangle$ }$\\
Then $\log P(A_{emo}=E|\tilde{U})$ is used for the SER reward. Additionally, the ASR reward is retained. After training, we compare the performance of the DiffRO-MTR model against the CosyVoice 2.0, GPT-SoVits, and F5-TTS, to evaluate their emotional expression capabilities. The test text and prompt audio also comes from CV3-Eval. Each emotion category consists of 100 samples, with Chinese and English evenly represented. We employ the emo2vec-plus-large model as the classifier and present the accuracy results in Table \ref{tab:ser_result}. 

\begin{table}[!t]
    \centering
    \begin{tabular}{l ccc ccc}
    \hline
        \multirow{2}{*}{Model} & \multicolumn{2}{c}{HAPPY}& \multicolumn{2}{c}{SAD} & \multicolumn{2}{c}{ANGRY}\\
        & zh & en & zh & en & zh & en  \\
    \hline
        F5-TTS\cite{chen-etal-2024-f5tts} &  0.92 & 0.64 & 0.68 &  0.92 & 0.40 & 0.76 \\
        GPT-SoVITS &  0.87 &  0.39 &  0.52 & 0.88 &  0.68 & 0.48 \\
        CosyVoice2.0 &  0.92 &  0.80 &  0.52 & 0.84 & 0.76 & 0.80  \\
         + DiffRO-MTR &  \textbf{1.00} &  \textbf{0.92} &  \textbf{0.76} & \textbf{0.96} & \textbf{0.84} & \textbf{0.92} \\

    \hline
    \end{tabular}
    \caption{Accuracy comparison for Audios synthesized from Different TTS system.}
    \label{tab:ser_result}
\end{table}

\begin{figure}
    \centering
    \includegraphics[width=1.0\linewidth]{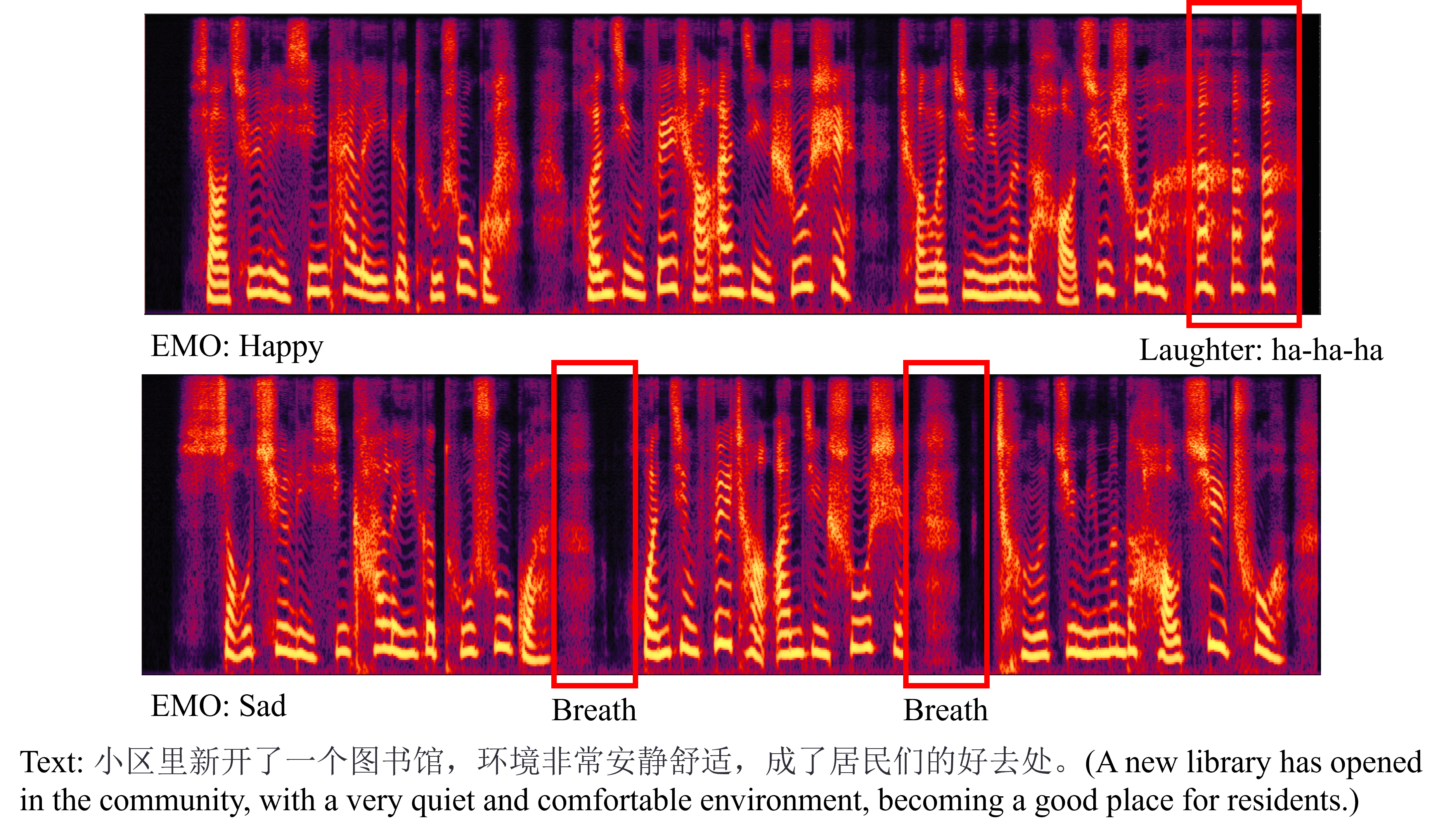}
    \caption{Laughter (top) and breath (down) generated in different emotion with same text.}
    \label{fig:event}
\end{figure}

The results demonstrate that the proposed DiffRO significantly enhances the emotional expressiveness of the TTS system. Notably, there are no emotion-labeled data available during the RL phase; in other words, emotional knowledge is acquired from the MTR model rather than from explicit data. Another intriguing finding is that the TTS system can learn to synthesize laughter, sobs, breaths, and other audio events to convey emotion. Examples of these findings are illustrated in Figure \ref{fig:event}.

\subsubsection{Other Attribute Control}

We also try to control other voice attributes such as MOS, age and gender. Experiments prove that DiffRO can influence the pronunciation, but, however, cannot play a decisive role. We show the results of the MOS control in Table \ref{tab:mos_result}, which uses $\log P(A_{sqa}=\text{MOS}_{t}|\tilde{U})$ as a reward. We can find the DiffRO do influence the audio quality, however, the MOS predicted from codec tokens can be close to the target, while the final generated audio will be higher (expected the $\text{MOS}_t=4$). 
Because the DiffRO is only applied on the LM, which can learn to generate noisy tokens in an effort to minimize the MOS. However, the ultimate quality of the audio output is more heavily influenced by the FM and the vocoder, both of which are trained on clean audio and possess denoising capabilities. This issue becomes particularly pronounced when controlling speaker attributes such as age and gender, especially since the speaker information is integrated during the FM phase in the CosyVoice 2.0 framework.

\begin{table}[!t]
    \centering
    \begin{tabular}{l ccc}
    \hline
             \multirow{2}{*}{Model} & \multicolumn{3}{c}{MOS}  \\
         & Codec & Audio & Recodec   \\
    \hline
        Baseline &  3.82 & 3.84 & 3.88 \\
        $\text{MOS}_{t}=2$ &  2.20 &  3.77 & 3.47 \\
        $\text{MOS}_{t}=3$ &  3.14 &  3.81 & 3.66 \\
        $\text{MOS}_{t}=4$ &  3.96 &  3.86 & 3.92 \\

    \hline
    \end{tabular}
    \caption{MOS score compassion for the TTS system. MOS-Codec is the average expectation MOS calculated from the predicted codec token $\tilde{U}$ by the MTR model. MOS-Audio is calculated from the synthesized speech $\tilde{X}$ by DNSMOS. We re-encode the $\tilde{X}$ into codec tokens and calculate the MOS-Recodec by the MTR model.}
    \label{tab:mos_result}
\end{table}

\section{Conclusions}
\label{S:5}
% In this paper, we propose a novel DiffRO based RL strategy for LLM-based TTS systems. Compared with other RL method, the DiffRO can directly predict the reward score from the speech token rather than the synthesized audios. And it can directly optimize the LM parameters by back-propagation. In addition, we develop an MTR model that can not only improve the pronunciation accuracy but also imporve the emotion expression and other sound attribute by different downstream task reward.
% For the further works, we will add more downstream tasks into the MTR model and try to apply the DiffRO on the FM module, which play more important roles in the speaker-related downstream tasks.

In this paper, we propose a novel DiffRO based reinforcement learning strategy for neural codec LM based text-to-speech systems. Compared to other reinforcement learning methods, DiffRO is capable of directly predicting reward scores from speech tokens rather than from synthesized audio. Moreover, it allows for the direct optimization of LM parameters through backpropagation.
Additionally, we have developed a MTR model that not only enhances pronunciation accuracy but also improves emotional expression and other acoustic attributes through various downstream task rewards.
For future work, we plan to incorporate additional downstream tasks into the MTR model and explore the application of DiffRO to the FM module, which plays a crucial role in speaker-related downstream tasks.

\bibliographystyle{IEEEtran}

\bibliography{mybib}

% \begin{thebibliography}{9}
% \bibitem[1]{Davis80-COP}
%   S.\ B.\ Davis and P.\ Mermelstein,
%   ``Comparison of parametric representation for monosyllabic word recognition in continuously spoken sentences,''
%   \textit{IEEE Transactions on Acoustics, Speech and Signal Processing}, vol.~28, no.~4, pp.~357--366, 1980.
% \bibitem[2]{Rabiner89-ATO}
%   L.\ R.\ Rabiner,
%   ``A tutorial on hidden Markov models and selected applications in speech recognition,''
%   \textit{Proceedings of the IEEE}, vol.~77, no.~2, pp.~257-286, 1989.
% \bibitem[3]{Hastie09-TEO}
%   T.\ Hastie, R.\ Tibshirani, and J.\ Friedman,
%   \textit{The Elements of Statistical Learning -- Data Mining, Inference, and Prediction}.
%   New York: Springer, 2009.
% \bibitem[4]{YourName17-XXX}
%   F.\ Lastname1, F.\ Lastname2, and F.\ Lastname3,
%   ``Title of your INTERSPEECH 2021 publication,''
%   in \textit{Interspeech 2021 -- 20\textsuperscript{th} Annual Conference of the International Speech Communication Association, September 15-19, Graz, Austria, Proceedings, Proceedings}, 2020, pp.~100--104.
% \end{thebibliography}

\end{document}